\documentstyle[preprint,aps,epsfig]{revtex}

\begin{document}


\title{Self-similarity and coarsening of three dimensional particles on
a one or two dimensional matrix}

\author{Jorge Vi\~nals}
\address{Supercomputer Computations Research Institute, 
Florida State University,
Tallahassee, Florida 32306-4052, and Department of Chemical Engineering,
FAMU/FSU College of Engineering, Tallahassee, Florida 32310-6046}
\author{W.W. Mullins}
\address{Department of Materials Science and Engineering, Carnegie 
Mellon University, Pittsburgh, Pennsylvania 15213-3890}

\date{\today}

\maketitle

\begin{abstract}

We examine the validity of the hypothesis of self-similarity in
systems coarsening under the driving force of interface energy reduction
in which three dimensional particles are
intersected by a one or two dimensional diffusion matrix.
In both cases, solute fluxes onto the surface of the
particles, assumed spherical, depend on both particle radius and
inter-particle distance. We argue that overall mass conservation
requires independent scalings for particle sizes and inter-particle 
distances under magnification of the structure, and predict power law growth
for the average particle size in the case of a one dimensional matrix (3D/1D), 
and a weak breakdown of self-similarity in the two dimensional case (3D/2D). 
Numerical calculations confirm our 
predictions regarding self-similarity and power law growth of average particle
size with an 
exponent 1/7 for the 3D/1D case, and provide evidence for the 
existence of logarithmic factors in the laws of boundary motion for 
the 3D/2D case. The latter indicate a weak breakdown of self-similarity.

\end{abstract}



\section{Introduction}
\label{sec:introduction}

The purpose of this paper is to re-examine the validity of the hypothesis of
self-similarity in coarsening systems in which particles of dimension
$D_{p}$ are
intersected by a diffusion matrix of different spatial dimensionality $D$. 
The discussion is limited to coarsening driven by interface energy reduction,
the particles are assumed to remain spherical (e.g., by surface diffusion), 
and particle migration is neglected.
We focus on two specific cases: three dimensional particles intersected by 
either a two (3D/2D) or one (3D/1D) dimensional diffusion matrix respectively. 

The statistical self-similarity hypothesis with one scaling length 
asserts that after a possible
transient, consecutive configurations of the coarsening structure are
geometrically similar in a statistical sense \cite{re:mullins86,re:mullins89}.
As a consequence, any parameter of the structure that is invariant under a
uniform magnification is also independent of time. This hypothesis, together
with the laws of boundary motion for a specific system (and their scaling
under uniform magnification) are sufficient to obtain the equation of motion
for any linear scale of the structure (e.g., the average particle radius
$\left< R(t) \right>$ for an ensemble of coarsening spherical domains). 

Self-similarity with a single scaling length as stated above is
consistent with conservation of mass (volume) only in systems for which 
$D_{p} = D$. Thus conservation of particle mass requires,
\begin{equation}
\label{eq:mcon}
n_{A} \left< R^{D_{p}} \right> = \frac{ \left< R^{D_{p}} \right> }
{d_{av}^{D}} = {\rm const.}
\end{equation}
where $n_{A}$ is the number of particles per unit (general) area of substrate
and $d_{av} = 1/n_{A}^{1/D}$ may be regarded as an average spacing between
particles. If self-similarity holds for the particle size distribution, then
\begin{equation}
\left< R^{D_{p}} \right> = {\rm const.} \left< R \right>^{D_{p}}
\end{equation}
so that Eq. (\ref{eq:mcon}) can be written
\begin{equation}
\label{eq:genscal}
\frac{ \left< R \right>^{D_{p}/D}}{d_{av}} = {\rm const.}
\end{equation}
Self-similarity with one scaling length requires $\left< R \right> /d_{av} 
= {\rm const}.$ But according to Eq. (\ref{eq:genscal}) this condition holds 
only if $D_{p} = D$. Therefore, self-similarity with a single scaling
length cannot hold when $D_{p} \neq D$. Unfortunately, self-similarity with one
scaling length was used in a section of Ref. \cite{re:mullins86} to deduce
erroneously exponents of particle size in a 3D/2D and a 3D/1D system ($t^{1/4}$
and $t^{1/5}$ respectively).

If $D_{p} \neq D$, self-similarity with two scaling lengths (e.g.,
$\left< R \right>$ and $d_{av}$) is still possible and
is consistent with Eq. (\ref{eq:genscal}). In this case, the
self-similarity hypothesis would assert that consecutive configurations of
the system are statistically equivalent to those obtained by magnifying
the original system using two scale factors, one for particle sizes
and a second for inter-particle spacings; the two scale factors are
related through Eq. (\ref{eq:genscal}) to conserve mass. We shall present
evidence that this occurs in the 3D/1D case and leads to a particle growth
exponent of 1/7. 

In the 3D/2D system, we will present numerical evidence that a 
necessary condition for self similarity is not met. This condition,
established in Appendix A, shows that if the distribution of reduced
particle sizes is scale invariant (which it must be in any self-similar
regime), then the ratio of the expected growth rates of any two particles 
must be scale invariant under magnification. We show numerically that this 
condition is not met in
the case of a configuration with a large number of interacting particles 
of unequal sizes, randomly distributed in space with a Laplacian
concentration field satisfying 
self-consistently determined mean field boundary conditions at large 
distances from the particles.

Support for the existence of self-similarity is quite strong in a large 
variety of physical systems for which $D_{p}=D$ (see, e.g., reviews in
\cite{re:gunton83,re:binder84b,re:furukawa85,re:bray94}).
In the case of three
dimensional particles growing by diffusion on a two dimensional substrate (the
3D/2D case), experiments and theoretical investigations have been recently
reviewed in \cite{re:zinke92}. Briefly, the system of coarsening particles
reaches (at least approximately) a scale invariant state in which the average
particle radius grows in time as $\left< R(t) \right> \propto t^{1/4}$
\cite{re:zinke95}. 
Such a growth law has been also predicted on the basis of mean field analyses
\cite{re:chakraverty67,re:speight68,re:kirchner71,re:gjostein75}.
We shall present evidence that self-similarity with this growth law
cannot be strictly true.

In Section \ref{sec:theory}, we use the law of boundary motion in a coarsening
system, together with self-similarity, where applicable, to discuss growth
exponents. In Section \ref{sec:3d1d}, we present numerical evidence that in the
3D/1D system, the particle radii and inter-particle distances scale separately
to obey Eq. (\ref{eq:genscal}) with $\left< R \right> \propto t^{1/7}$ and
$d_{av} \propto t^{3/7}$. Section \ref{sec:3d2d} addresses the 3D/2D case and
presents numerical evidence for the existence of a logarithmic
factor in the laws of boundary motion that involves ratios of particle sizes to
inter-particle distances, and indirect evidence that, 
contrary to the classical case of coarsening of two dimensional particles on a
two dimensional substrate, the logarithmic factor leads to a (weak) 
breakdown of self-similarity.
This breakdown would manifest itself in the existence of effective coarsening
exponents that change very slowly in time.
 
\section{Self-similarity and growth exponents}
\label{sec:theory}

In this section we present the derivation of the growth exponent for the case
$D_{p} =3, D=1$ and discuss the complication that arises in the case of
$D_{p}=3, D=2$. The discussion is based on the following theorem: If the
reduced particle size distribution is independent of scale, that is, if
\begin{equation}
\label{eq:ab}
n(R,t) = \frac{N(t)}{\left< R \right>} P(R/\left< R \right> ),
\end{equation}
where $n(R,t)dR$ is the number of particles per unit \lq\lq area" in the
system with (volume equivalent) radii in $dR$, $\left< R \right>$ is the 
average particle size and $P(x)$, in which
$x=R/\left< R \right>$, is the reduced particle size distribution 
function, then it is shown in the Appendix that
\begin{equation}
\label{eq:aj}
\left< \dot{R} | R \right> = \frac{d \left< R \right> }{dt} G(x),
\end{equation}
where $ \left< \dot{R} | R \right> $ is the expected value of $dR/dt$ for
particles of radius $R$. The expression for $G(x)$ given in the appendix shows 
that as $x$ increases from zero, $G$ increases from negative values, 
corresponding to particles that shrink on the average, to positive values
corresponding to particles that grow on the average; $G$ vanishes at a
value $x=x_{c}$, corresponding to particles that, on the average, do not
change size. Eq. (\ref{eq:aj}) shows that the ratio of the expected value of
$dR/dt$ for particles of two different sizes is independent of scale as
stated in the introduction. We will use this property (which we will
refer to as the ratio test) to argue that the 3D/2D system cannot be
strictly scale invariant based on the numerical evidence presented in
Section \ref{sec:3d2d}.

If self-similarity holds (with one or more scaling lengths), the growth
exponent may be determined from the scaling of $d \left< R \right> /dt$.
But Eq. (\ref{eq:aj}) shows that $d \left< R \right> /dt$ scales as
$\left< \dot{R} | R \right>$ which in turn scales as $dR/dt$ for a
single particle. Hence the growth exponent may be determined from the
scaling of $dR/dt$.

To discuss the scaling of $dR/dt$, we consider for simplicity an
ensemble of precipitate particles embedded in a matrix, such that their
growth or dissolution is limited by diffusion of one of the species
(solute) through the matrix (the discussion can be easily generalized to
other systems in which coarsening is driven by surface free energy
reduction \cite{re:mullins86,re:mullins89}). Let $c$ be the
concentration of solute in the matrix, and assume that $c$ in the matrix
is much smaller than the (constant) concentration in the precipitate
phase $c_{p}$, so that $c_{p} - c$ can be approximated by $c_{p}$ 
\cite{re:marqusee83}. Conservation
of solute mass requires that for each precipitate particle,
\begin{equation}
\label{eq:flux_law}
S_{D_{p}} \frac{dR}{dt} = - \frac{1}{c_{p}} \int j_{n} dS,
\end{equation}
where
$j_{n}$ is the normal component of the solute flux along the outward normal to 
the surface, $S_{D_{p}}$ is the surface of a particle of radius $R$ in
$D_{p}$ dimensions and is, for example, $S_{D_{p}} = 
\pi^{D_{p}/2} R^{D_{p}-1} / \Gamma(D_{p}/2)$ for a hemisphere,
and the integral is taken over the surface of the precipitate
particle. 

For the 3D/1D case, the collection area over which $j_{n}$ is nonzero
is of microscopic size and independent of scale. Let all particle radii
be multiplied by $\lambda$. Then the
flux $j_{n}$ scales as a gradient which scales as $1/(\lambda d_{av})$, 
the factor $\lambda$ arising from the Gibbs-Thomson equation. But, from Eq.
(\ref{eq:genscal})
$\lambda^{3}/d_{av} = {\rm const.}$ Hence Eq. (\ref{eq:flux_law}) shows that
$\left< \dot{R} | R \right>$ scales as $\lambda^{-6}$.
Therefore, if self-similarity holds, Eq. (\ref{eq:aj}) shows that
$d \left< R \right> /dt$ scales as $\lambda^{-6}$, or as $\left< R
\right>^{-6}$ so that
\begin{equation}
\frac{ d \left< R \right> }{dt} = {\rm const.} \left< R \right>^{-6},
\end{equation}
which integrates to $\left< R \right> \propto t^{1/7}$ and
$d_{av} \propto t^{3/7}$. Evidence supporting this result in presented in
Section \ref{sec:3d1d}.

To discuss the case of a two dimensional matrix we first present an 
expression for $dR/dt$
in the spirit of a mean field treatment. Consider a disk of
radius R located at the origin, and assume quasi-steady diffusion in the
matrix ($\nabla^{2} c = 0$), with boundary conditions $c(r=\xi) = c_{\xi}$
at some cut-off distance away from the center, and
\begin{equation}
\label{eq:gt}
c(r=R) = c_{0} \left( 1 + \frac{\Gamma}{R} \right),
\end{equation}
at the disk's boundary, where $c_{0}$ is the solute concentration in the
matrix at coexistence, and $\Gamma$ is the capillary length. Then the
rate of change of volume of the particle is proportional to the gradient
of solute at the disk times the particle perimeter or
\begin{equation}
\label{eq:am}
S_{D_{p}} \frac{dR}{dt} = - \frac{1}{c_{p}} \int j_{n} dS =
\frac{2 \pi D_{c} c_{0} \Gamma a_{0} }{c_{p} \ln (\xi /R) } \left(
\frac{1}{R_{c}} - \frac{1}{R} \right),
\end{equation}
where the concentration $c_{\xi}=c_{0}(1+\Gamma/R_{c})$ at the cutoff distance 
is set so that $dR/dt=0$ for $R=R_{c}$, $D_{c}$ is the solute diffusivity in
the matrix and $a_{0}$ is a microscopic length that defines the width of
the particle's collection area. For the 2D/2D case, self-similarity with one
scaling length is consistent with Eq. (\ref{eq:am}) since $\xi /R$ does not
change, and hence the ratio of $dR/dt$ for any two particles as given by Eq. 
(\ref{eq:am}) is independent of scale as required by Eq. (\ref{eq:aj}).
The result $ \left< R \right> \propto t^{1/3}$ then follows 
from Eq. (\ref{eq:am}), self-similarity and Eq. (\ref{eq:aj}).

In the 3D/2D case, self similarity with one scaling length would require
$\xi$ to scale as $R$ in Eq. (\ref{eq:am}). This was assumed by Chakraverty in 
a mean field treatment with the result $ \left< R \right> \propto t^{1/4}$
which again follows from
Eqs. (\ref{eq:am}) and (\ref{eq:aj}), and self-similarity. If, however, $\xi$ 
scales as $d_{av}$, the
inter-particle separation, then it must scale in accord with Eq. 
(\ref{eq:genscal}) and therefore
$\xi /R$ is no longer independent of scale. It would then follow from Eq. 
(\ref{eq:am})
that Eq. (\ref{eq:aj}) cannot hold, and the particle distribution function 
cannot be of the form (\ref{eq:ab}). We present evidence in Section
\ref{sec:3d2d} that this is the case, independent of the mean
field approximation, and hence that
self-similarity cannot hold strictly. We point out, however, that
the argument of the logarithm is expected to change slowly in time,
and hence the logarithmic factor itself will be changing slowly. Hence,
approximate scale invariance could be expected over relatively long
periods of time.

\section{Three dimensional particles on a one dimensional matrix}
\label{sec:3d1d}

We present in this section the numerical solution of a model system
comprising spherical three dimensional particles arranged on a closed 
loop (i.e., 
a line with periodic boundary conditions at the ends), such 
that each particle can only exchange mass with its two nearest neighbors.
Such a configuration is intended to model diffusion controlled coarsening of
three dimensional particles when transport through the matrix takes place
preferentially along a line (e.g., along a dislocation line).
The model, originally introduced by Hunderi {\it et al.} \cite{re:hunderi78}, 
is a one dimensional version of the so-called bubble models of grain growth in
polycrystalline materials. A mean field solution has shown that coarsening 
proceeds in a self-similar fashion, and the growth law for
the average particle size has been calculated \cite{re:mullins91}. 
The one dimensional model has been extended to study the existence of 
self-similarity when
multiple grain orientations and grain boundary anisotropies are allowed
\cite{re:mullins93}.

We consider a set of $N$ spherical particles of radii $R_{i}, i=1, \ldots, N$, 
forming a
linear chain with periodic boundary conditions. Particles are arranged on an
evenly spaced grid, such that the initial inter-particle separation is one. The
rate of change of each particle radius is given by
\begin{equation}
\label{eq:lbm}
R_{i}^{2} \frac{dR_{i}}{dt} = \frac{M}{d_{i,i+1}} \left( \frac{1}{R_{i+1}} - 
\frac{1}{R_{i}} \right) + \frac{M}{d_{i,i-1}} \left( \frac{1}{R_{i-1}} -
\frac{1}{R_{i}} \right),
\end{equation}
where $M$ is a mobility coefficient that sets the time scale appropriate for 
the microscopic mechanism responsible for diffusion, and $d_{i,i+1}$ 
and $d_{i,i-1}$ are the 
distances between particles $i$ and its two nearest neighbors $i+1$ and 
$i-1$ respectively.
As the system evolves, some particles shrink to zero radius and are 
removed. Therefore the total number of particles $N$ decreases with
time whereas both average particle size
and average inter-particle distance increase. It immediately 
follows from Eq. (\ref{eq:lbm}) that the total volume of the
set of particles, $V = (4 \pi/3) \sum_{i=1}^{N} R_{i}^{3}$ 
is independent of time.

We initially place a large number of particles ($N = 3 \times 10^{6}$) on 
a line, and impose periodic boundary 
conditions $R_{N+1}(t) = R_{1}(t)$, and $d_{N,1}(t) = d_{1,N}(t)$.
The Euler method is used to integrate the system of 
equations (\ref{eq:lbm}) with $M = 1$ and a step size $\Delta t = 5$.
The initial condition is a set of randomly chosen radii, uniformly distributed
between $R_{min}$ and $R_{max}$. The value of $R_{min}$ is chosen so that
the algorithm is stable, and that no particle with $R > R_{min}$ can 
shrink to 
zero radius in $\Delta t$. We have further chosen $R_{max} = 10$ arbitrarily.
The numerical solution proceeds as follows. Given a configuration at 
time $t$,
$\{ R_{i}(t) \}, i=1, \ldots , N(t)$, Eq. (\ref{eq:lbm}) is iterated once 
for each particle to yield
$ \{ R_{i} (t+\Delta t) \}$. Any particle for which $R(t+\Delta t) <
R_{min}$ is eliminated, so that only $N(t+ \Delta t)$ particles remain.
Links are then redefined so that each particle is connected to its two nearest
neighbors but preserving their original relative distances.
We have studied 420,000 iterations, at the end of which 99053 particles 
were left.
We check the accuracy of the integration by monitoring the total volume
of the set of particles. At very early times, a large number of particles are 
lost and the volume after the first 10,000 iterations decreased by 2.5 \%.
From then on to the end of the calculation, the total volume only changed 
by 0.3 \%.

\begin{figure}[t]
\psfig{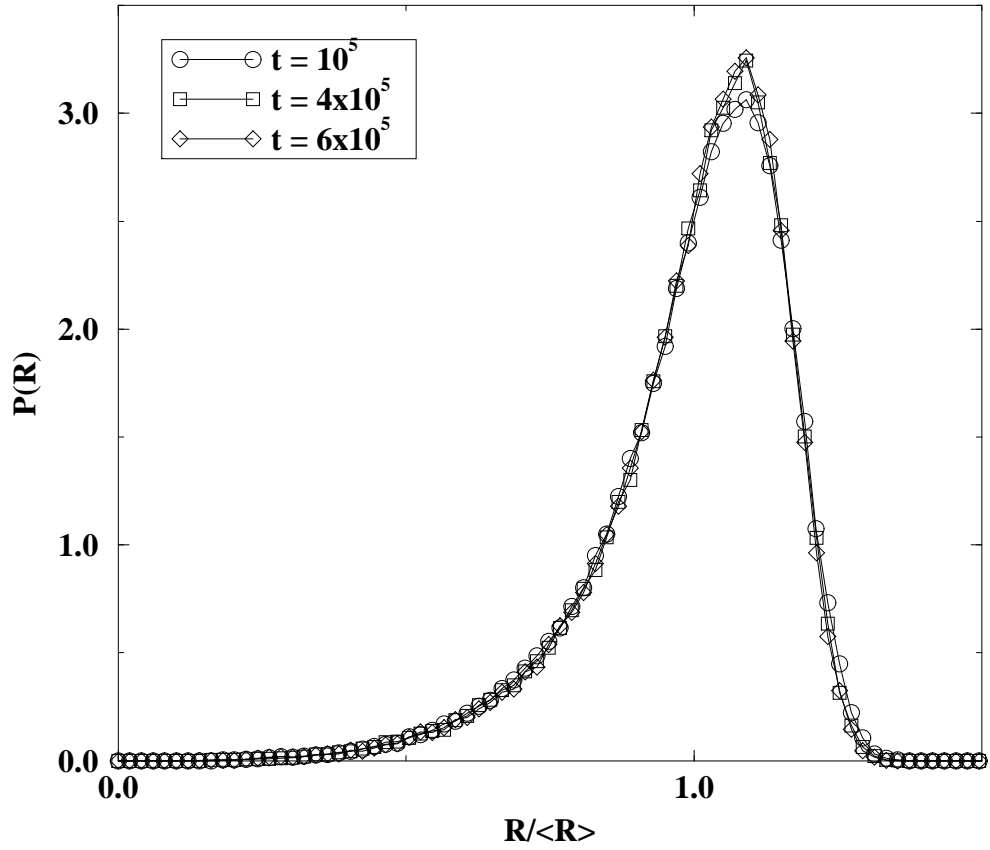}
\caption{Scaled distribution of particle radii at three different times as 
indicated, for three dimensional spherical particles on a line.
At the earliest time the system is still approaching a scale invariant state.}
\label{fi:lbmpr}
\end{figure}

\begin{figure}[t]
\psfig{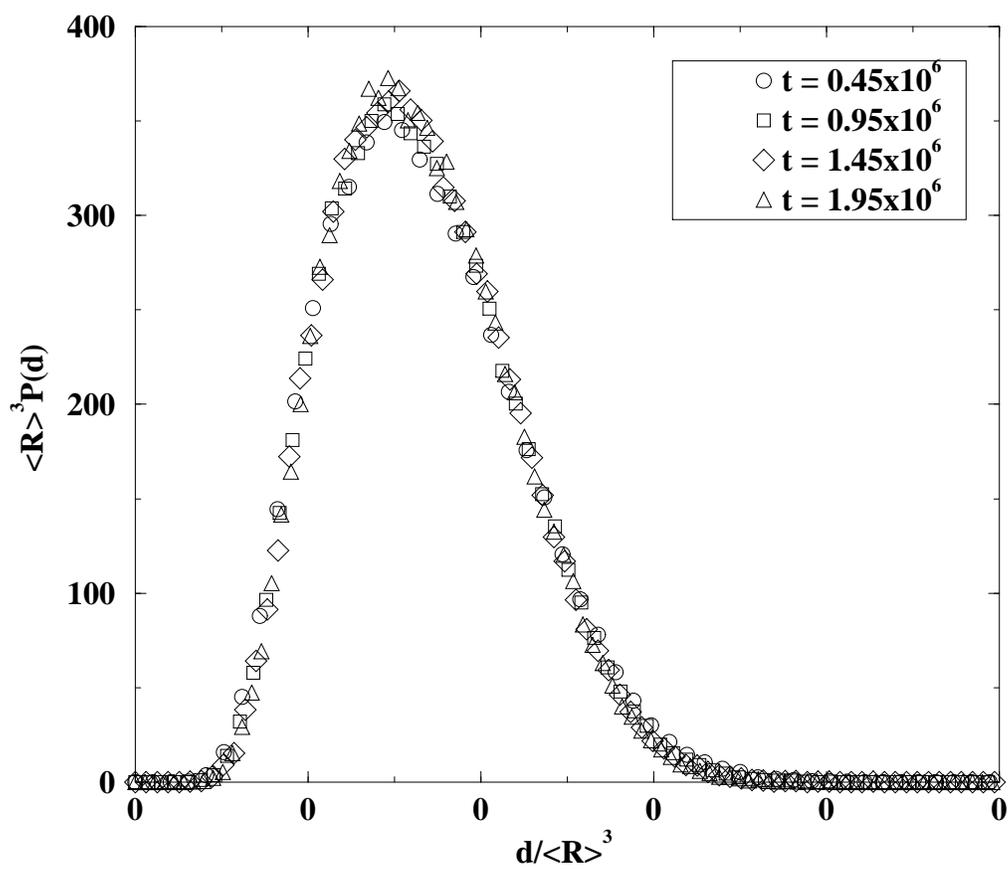}
\caption{Distribution of inter-particle separations, $P(d)$, scaled by the
average particle size $\left< R \right>^{3}$ for the times shown.}
\label{fi:lbmpd}
\end{figure}

Figure \ref{fi:lbmpr} shows the scaled distribution of particle radii at three 
different times.
The earlier time is slightly before entering the self-similar regime, the other
two are arbitrary times within it. All other later times agree with these 
two within 
the size of the symbols, indicating the existence of a self-similar regime. 
In addition, $\left< R \right>^{3} / <d> \simeq 0.0033$ 
($\left< d \right> = d_{av}$)
changes by less that 0.2 \% between $t = 10^{5}$ and the end of the 
calculation, in agreement with the two-length scaling
presented in Section \ref{sec:introduction} (Eq. (\ref{eq:genscal})).
Figure \ref{fi:lbmpd} further shows the distribution of inter-particle 
distances scaled with the average particle radius to the third power, 
also consistent with the predictions of Section 
\ref{sec:introduction} with regard to the existence of two scaling lengths.

\begin{figure}[t]
\psfig{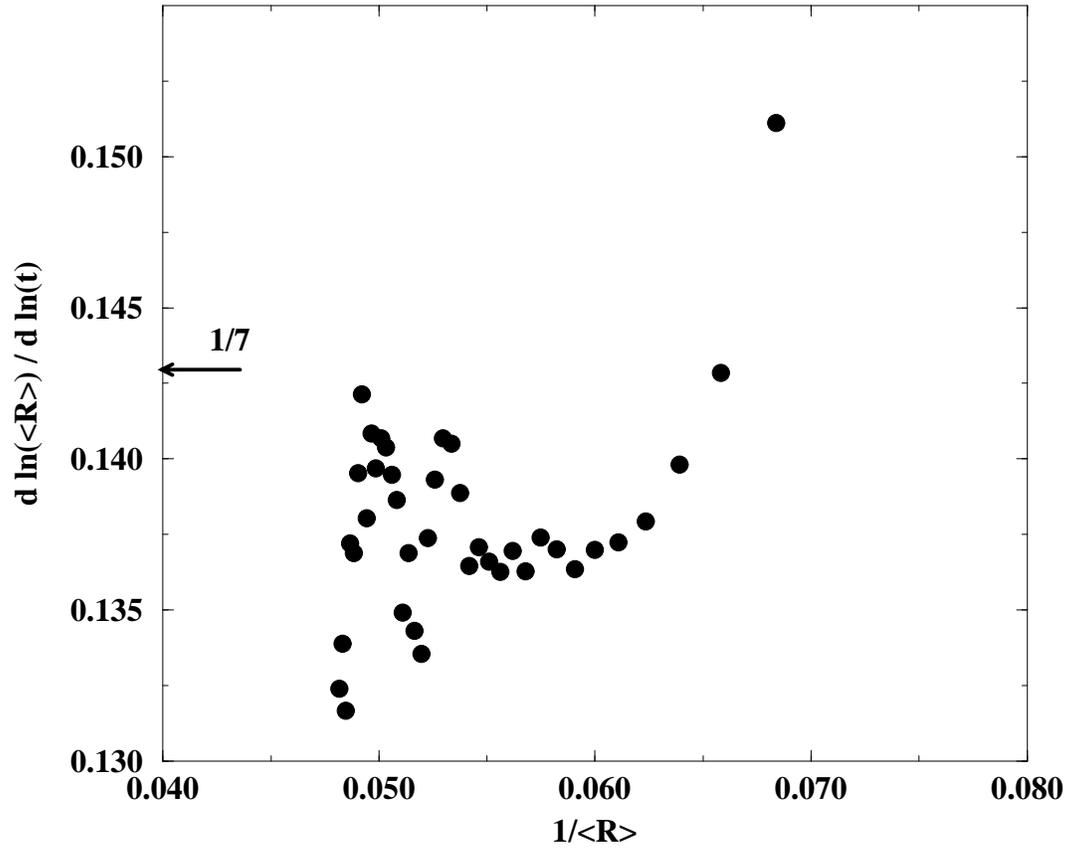}
\caption{Logarithmic time derivative of the average particle radius 
$\left< R \right> (t)$ plotted versus $1/\left< R \right>$. 
For power law growth, the 
logarithmic derivative asymptotes to a constant equal to the value of the 
exponent $n$. The arrow on the left axis indicates the exact position of the 
value 1/7.}
\label{fi:lbmexp}
\end{figure}

In summary, even though this model has two independent length scales,
namely the characteristic particle size and inter-particle distance, there is 
an attractor
for the evolution in which each distribution function is scale invariant,
and each length scale satisfies a well defined relationship in the 
asymptotic limit of long times (although not proportional to each other).
Finally, Fig. \ref{fi:lbmexp} shows our best estimate of the exponent
$n$, which is very close to the theoretical prediction of $n = 1/7$.

\section{Three dimensional particles on a two dimensional matrix}
\label{sec:3d2d}

Self-similarity and the associated 
growth exponent are considerably more difficult to investigate numerically in 
this case. Previous numerical approaches in three dimensions
have considered configurations comprising a large number of 
point particles and computed growth rates by direct summation 
of the Green's function of the Laplace operator 
\cite{re:voorhees84,re:voorhees85,re:yao93}. 
We have not used this method because the calculation of the 
Green's function for
given boundary conditions requires infinite sums over the appropriate images, a
procedure that does not converge in two dimensions.
Furthermore, possible deviations from self-similarity depend logarithmic on 
time, and hence are very weak. As a consequence, numerical integration
of some particular model would require long time spans to unambiguously 
discriminate between such a dependence and power law growth with some 
effective exponent. We have therefore focused on the ratio test, namely
on whether the ratio of expected growth rates of particles of different
sizes is independent of scale. If the ratio is not independent of scale,
then the distribution of particle sizes cannot be scale invariant (i.e.,
Eq. (\ref{eq:ab}) cannot hold). The generality of this procedure
is limited by the specific particle distributions
that we use to apply the ratio test. These distributions are not obtained 
self-consistently by direct solution of a coarsening system.
It is therefore conceivable that the failure we find of the ratio test
for our distributions might not hold for a
coarsening system. However, all our numerical evidence clearly points to
logarithmic factors in the law of boundary motion for a two dimensional
matrix, and we think it is very unlikely that these factors would cancel for
a special class of configurations.

We discuss two cases. First, we present a numerical 
solution to Laplace's equation in a 
two dimensional square domain with circular disks of fixed radius and
concentration at the corners and periodic boundary
conditions; the configuration is equivalent to an infinite square
lattice with alternating small and big circles at the lattice sites.  The
solution establishes that, for this simplified configuration
of only two types of interacting particles, concentration fields in the
domain do have a logarithmic factor involving ratios of particle sizes to
inter-particle distances. Second, we extend this solution to an ensemble
of small discs with a far field boundary condition of mean-field type, 
which is also solved for self-consistently. The same conclusion holds
for this analysis. 

The results will be analyzed in terms of either $ \left< \dot{R} | R \right> $
directly, or in terms of integrated fluxes to particles which are easier
to determine numerically. For the latter purpose we define $Q$ to be the
integrated volume flux transferred
from all shrinking particles to all growing particles, and $q = Q/N$;
other related quantities are defined in section \ref{sec:lmf}. 
By definition, $Q = \int_{R_{c}}^{\infty} S_{D_{p}} J(R,t) dR$
(see Eq. (\ref{eq:flux_law}) for the definition of $S_{D_{p}}$). Given 
the definition of $J$ given in Appendix A (Eqs. (\ref{eq:apai})
and (\ref{eq:apabb})), and the
result (\ref{eq:aj}), one has, in the self-similar regime,
\begin{equation}
\label{eq:qdef}
q \propto \left< R \right>^{D_{p}-1} \frac{ d \left< R \right> }{dt};
\end{equation}
$q$ is constant in the standard 3D/3D case, and
inversely proportional to the average particle radius for 2D/2D.

In general, $q$ is a function of the radius and location of the centers
of all the particles, and of the position of the center of the outer
boundary $x_{0}$ (assumed, for example, spherical) and its radius $R_{0}$: 
$q = q(R_{0}, R_{i}; x_{0}, x_{j})$. The function $q$ is a
homogeneous function of degree -1 for a two dimensional substrate
allowing for capillarity. This follows from the observation that if
$c(\vec{r})$ is a solution of Laplace's equation for the original configuration
satisfying all boundary conditions, then $(1/\lambda ) c(\vec{r}/\lambda)$
is a solution for the configuration scaled up uniformly by
$\lambda$. Hence all gradients scale as $1/\lambda^{2}$, and integrated fluxes 
to each particle and therefore $q$ by $1/\lambda$. Now, starting with
a given spatial configuration, consider scaling up all particle radii by a 
factor $s$, and all centers of particle positions by a different factor
$t$. Then,
\begin{equation}
\label{eq:fts}
q(sR_{i}; tx_{j}) = q\left( sR_{i}; s\frac{t}{s}x_{j} \right) = 
\frac{1}{s} q \left(R_{i}; \frac{t}{s} \; x_{j} \right) =
\frac{1}{s} f(t/s),
\end{equation}
where $f(1) = q (R_{i};x_{j})$, and,
for simplicity, we have taken the origin of the outer boundary to
lie at $x_{0} = 0$. The function $f$ depends on the ratio of $s$
and $t$ only. As a consequence, it is sufficient to consider rescaling
particle distances at fixed particle radii ($s=1$) to obtain the scaling of
$q$, and, if self similarity holds, to
obtain the growth law for the average particle size through Eq. 
(\ref{eq:qdef}).

\subsection{Two interacting particles in a Laplacian field}
\label{sec:3d2d-2s}

\begin{figure}[t]
\psfig{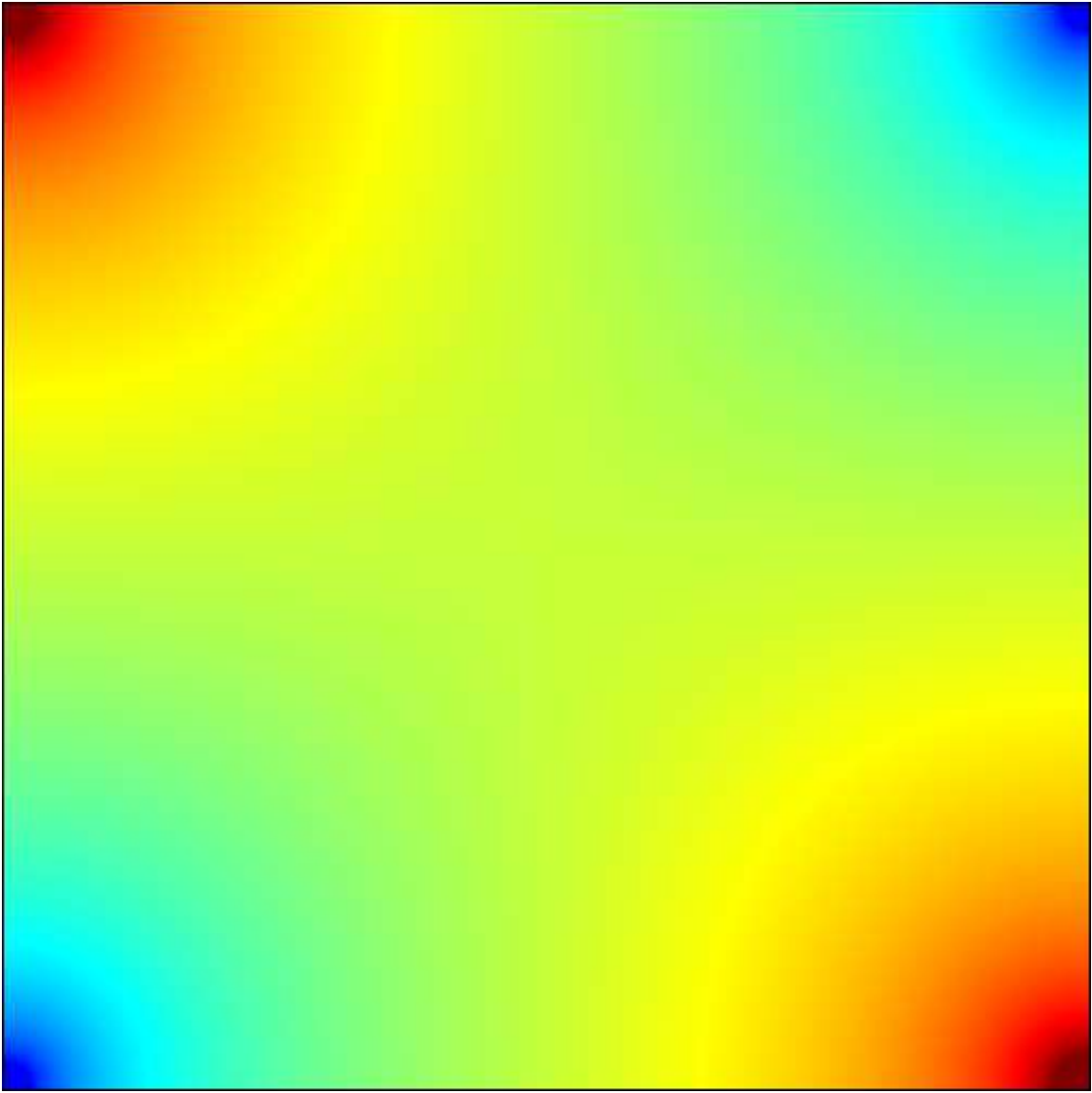}
\vspace{0.5cm}
\caption{Two dimensional configuration used in the solution 
of $\nabla^{2} c =0$; the side of the computational cell is $L=100$. 
Two discs 
of radius $R_{a}=3$ (left bottom) and $R_{b}=2$ (right bottom) are placed
at the corners of the square domain with periodic boundary conditions. The 
concentration field satisfies Laplace's equation outside the 
discs, and is constant inside and equal to $c_{a}=1$ and $c_{b}=-1$. 
We show in grey scale the iso-concentration lines.}
\label{fi:two-discs}
\end{figure}

The first case considered is shown in Fig.  \ref{fi:two-discs}. Two discs of 
radii $R_{a}$ and $R_{b}$, and concentration $c_{a}$ and 
$c_{b}$, are placed on the corners of a square lattice of side $L$ with
periodic boundary conditions. We have considered the case $c_{a} =1, R_{a} =
3$ and $c_{b} = -1, R_{b}=2$ with values of $L$ ranging from $L=50$ to
$L=500$. In order to model quasi-static diffusive transport in the matrix
Laplace's equation is solved in the interior region subject to the boundary
conditions specified at the discs boundaries and periodic boundary conditions
otherwise. The computational domain is discretized in $N_{p}$ evenly spaced
elements in each direction. The values of $N_{p}$ are adjusted for each $L$ in
order to obtain a reasonably accurate solution, and range from $N_{p}=500$ to
$N_{p}= 2500$. Laplace's equation for the concentration field has been solved
with a Successive Over-relaxation (SOR) method with Chebyshev acceleration
\cite{re:press86}. 

\begin{figure}[t]
\psfig{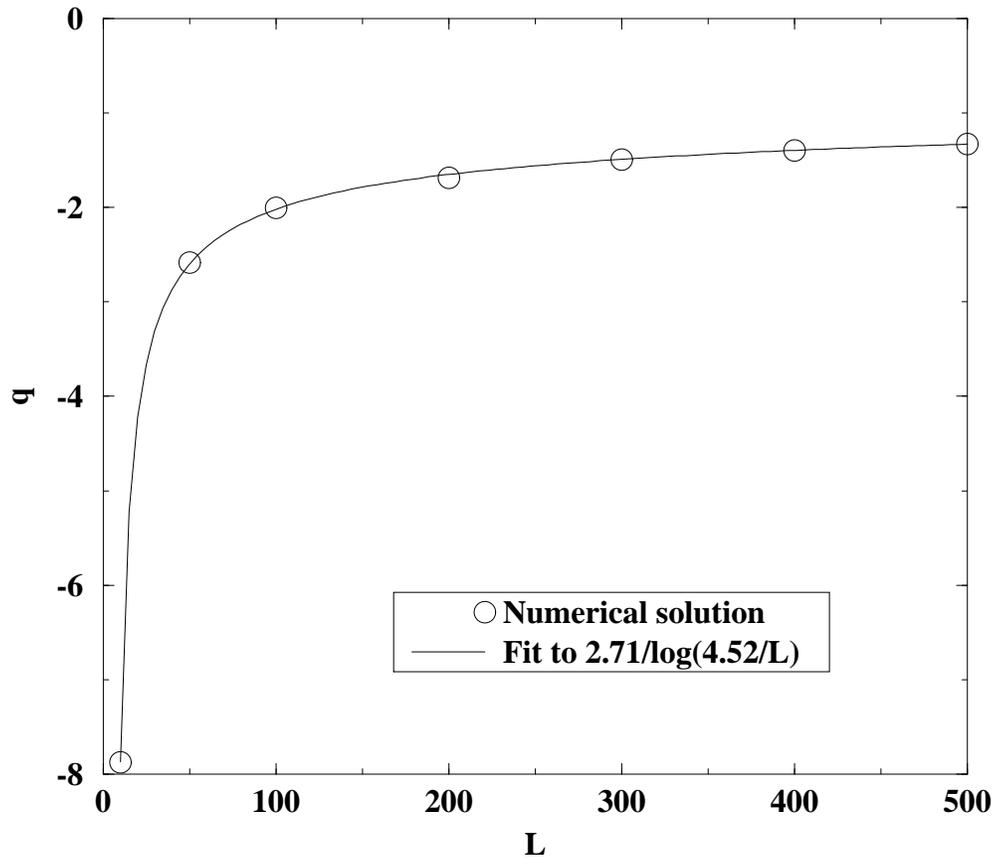}
\caption{Average rate of volume transfer from particle $b$ to particle $a$
as a function of the system size $L$.
The remaining parameters have been kept constant and equal
$R_{a} = 3, R_{b} = 2, c_{a} = 1$ and $c_{b} = -1$. The numerical
solution is indicated by the circles, and the solid line is a fit to the
logarithmic function shown. 
}
\label{fi:gradL}
\end{figure}

Figure \ref{fi:gradL} shows the dependence of $q$ with $L$ at constant
radii and concentration of the particles obtained from the direct solution to
Laplace's equation in the square domain.
The fit indicates the presence of a logarithmic factor in $q$ in the
solution, in agreement with the results of the mean field calculation
presented in Section \ref{sec:theory}. It shows, for this simple case,
that the function $f$ in Eq. (\ref{eq:fts}) has an approximate logarithmic 
dependence on its argument $t/s$. When combined with Eq. (\ref{eq:qdef}), 
the result indicates that the coarsening exponent cannot be strictly 1/4.
Furthermore, since $L$ is proportional to $d_{av}$, the inter-particle
spacing, the results indicate that $\xi$ in the mean field expression
(Eq. (\ref{eq:am})) is not proportional to $R$ but rather
(approximately) to $d_{av}$. If this is so, it follows that the ratio test
cannot be satisfied, the reduced particle size is not scale invariant
and strict self-similarity does not hold.

\subsection{Interacting particles in a Laplacian field}
\label{sec:lmf}

To investigate the possibility that the growth rates of individual
particles all contain the same logarithmic factor which therefore
cancels out when ratios are formed and allows the ratio test to be
satisfied, we have extended the calculations to a fairly large
collection of small interacting particles. The results reported below
again indicate that the ratio test fails and therefore that strict
self-similarity does not hold.

For numerical reasons, we have 
chosen elemental square particles of random side ranging from the grid 
spacing $h$ to $5h$. This eliminates the need to resolve circular contours, 
but introduces additional anisotropies (and singularities) into the solution. 
However, we have verified that both effects are negligible in the range of 
parameters used in our calculations. Furthermore, $c(r)$
becomes spatially isotropic at distances $\approx 3h$ from the square.

\begin{figure}[t]
\psfig{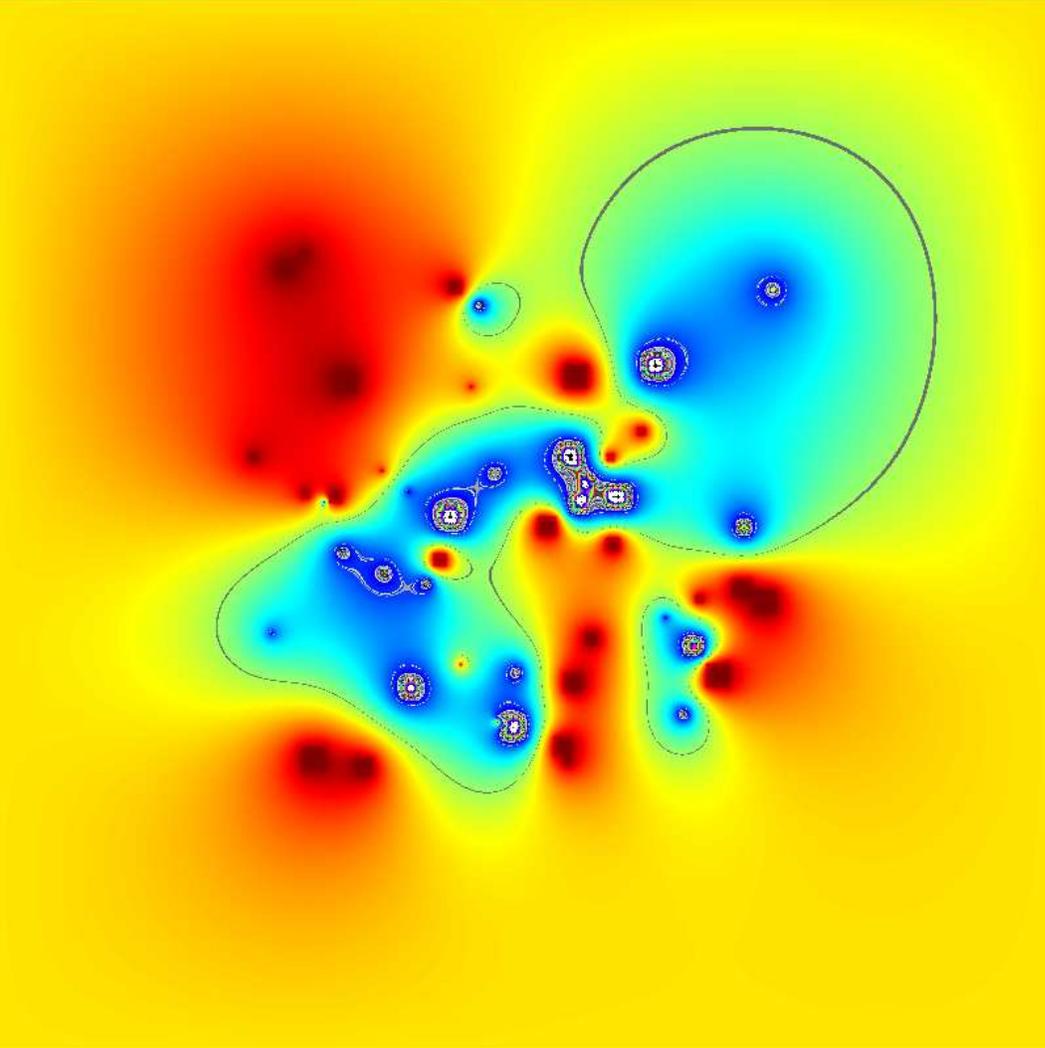}
\vspace{0.5cm}
\caption{Example of the configuration used in the numerical solution of
$\nabla^{2} c = 0$ subject to self-consistent mean field boundary
conditions at the outer boundary. Shown are 50 elemental domains
of random size (uniformly distributed between $h$ and $5h$, with $h$ the
lattice spacing). The composition of each domain is chosen randomly
as $\pm 1$ with equal probability. The 50 small
elements have been placed at random (uniformly distributed)
within the inner quarter of the computational domain.
We fix $c_{\infty}$ at the outer boundary ($x =0, x=L, y=0$ and $y = L$)
and compute the total flux through this boundary. As
described in the text, $c_{\infty}$ is then adjusted until
the total flux vanishes. The solution found is shown in the figure 
in grey scale. }
\label{fi:conf_mf}
\end{figure}

A large collection of such squares ($N = 50$) has been placed at random 
within the inner fourth of a square lattice of side $L$. We have randomly 
assigned the value $c_{p}= +1$ and $c_{p}=-1$ with equal probability (see Fig.
\ref{fi:conf_mf}). Laplace's equation for the concentration
is solved in the outer region with the
boundary condition $c=c_{\infty}$ (uniform) on the outer boundary. 
The total flux through the outer boundary $J_{\infty} = - \int \hat{n} \cdot
\nabla c dl$ is then computed. For an arbitrary
choice of $c_{\infty}$, $J_{\infty} \neq 0$. An iterative procedure is then
performed by adjusting the value of $c_{\infty}$ until $J_{\infty} = 0$. 
Such a procedure is intended to model a system of $N$ interacting precipitate 
particles embedded in a Laplacian field, chosen, as usual, so that 
the total mass is conserved.

\begin{figure}[t]
\psfig{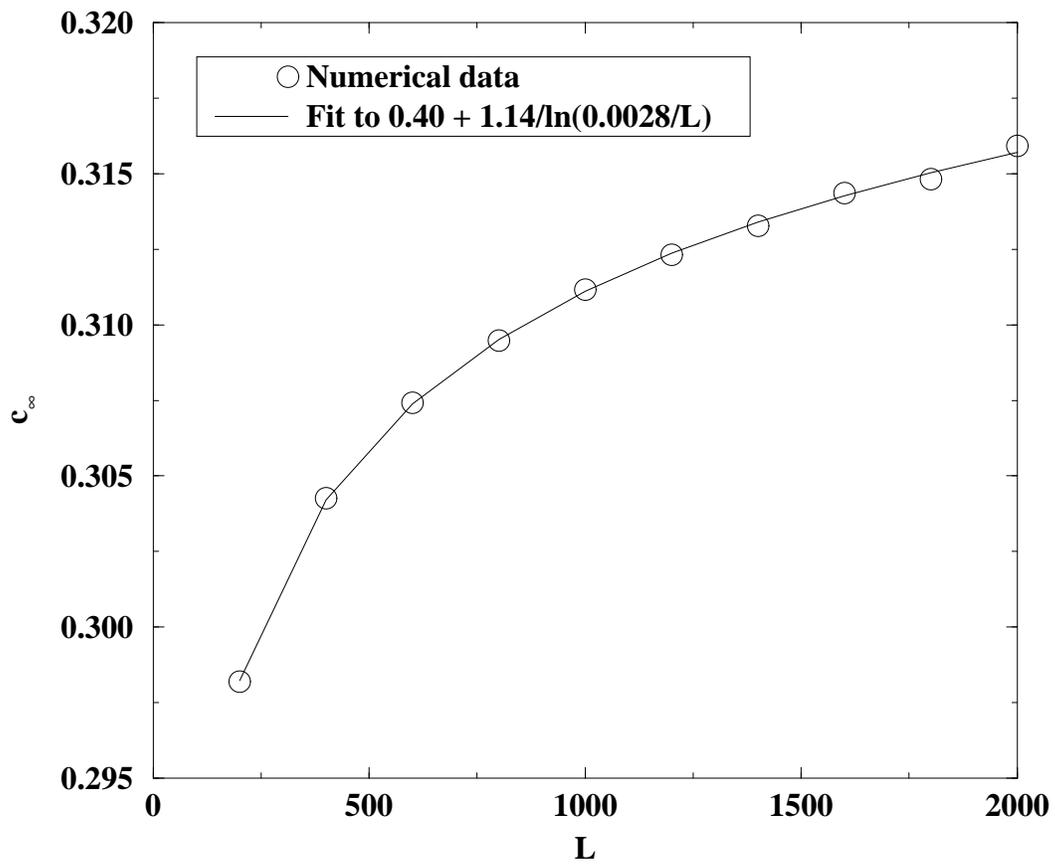}
\caption{Mean field concentration $c_{\infty}$ as a function of system size
$L$ for the configuration shown in Fig. \protect\ref{fi:conf_mf}.}
\label{fi:cinf}
\end{figure}

Once a self-consistent solution for a given $L$ has been found, the linear 
dimensions of the system are scaled up while keeping $h$ and $c_{p}$ constants
and the entire procedure is repeated. 
Figure \ref{fi:cinf} shows the value of $c_{\infty}$ as a 
function of $L$ for the configuration shown in Fig. \ref{fi:conf_mf}. 
There is an evident dependence of $c_{\infty}$ on $\ln L$.
The three parameter fit is needed to account for a non-zero value of the 
average concentration. We have also performed a similar numerical computation
involving discs of
the same size and fixed concentration (so that the average concentration can
be set to zero). A two parameter fit to the resulting function $c_{\infty}$ 
vs. $L$ is equally good.

We have repeated all our calculations for the related case that involves a 
large system with periodic boundary conditions, with the elemental squares 
being uniformly distributed throughout the entire computational domain. 
Identical results have been obtained, the details of which
will not be presented here.

We apply the ratio test in the following form: Eq. (\ref{eq:aj}) implies
that if $\alpha$ and $\beta$ are any two sets of particles each with
specified radii, then the ratio $\left< Q_{\alpha} \right> / \left< Q_{\beta}
\right>$ is independent of scale, where $\left< Q_{\alpha} \right>$ is
the expected value of the integrated flux to particles in set $\alpha$,
and similarly for $\left< Q_{\beta} \right>$. Thus
\begin{equation}
\left< Q_{\alpha} \right> = \sum_{i \in \alpha} 2 \pi R_{i}^{2} \left<
\dot{R}_{i} | R_{i} \right> = 2 \pi \left< R \right>^{2} \frac{ d \left<
R \right> }{dt} \sum_{i \in \alpha} x_{i}^{2} G(x_{i})
\end{equation}
and similarly for $\left< Q_{\beta} \right>$ so that in the ratio, the
prefactors to the sum cancel and the sums are independent of scale
($x_{i}$ here is the reduced particle radius).

\begin{figure}[t]
\psfig{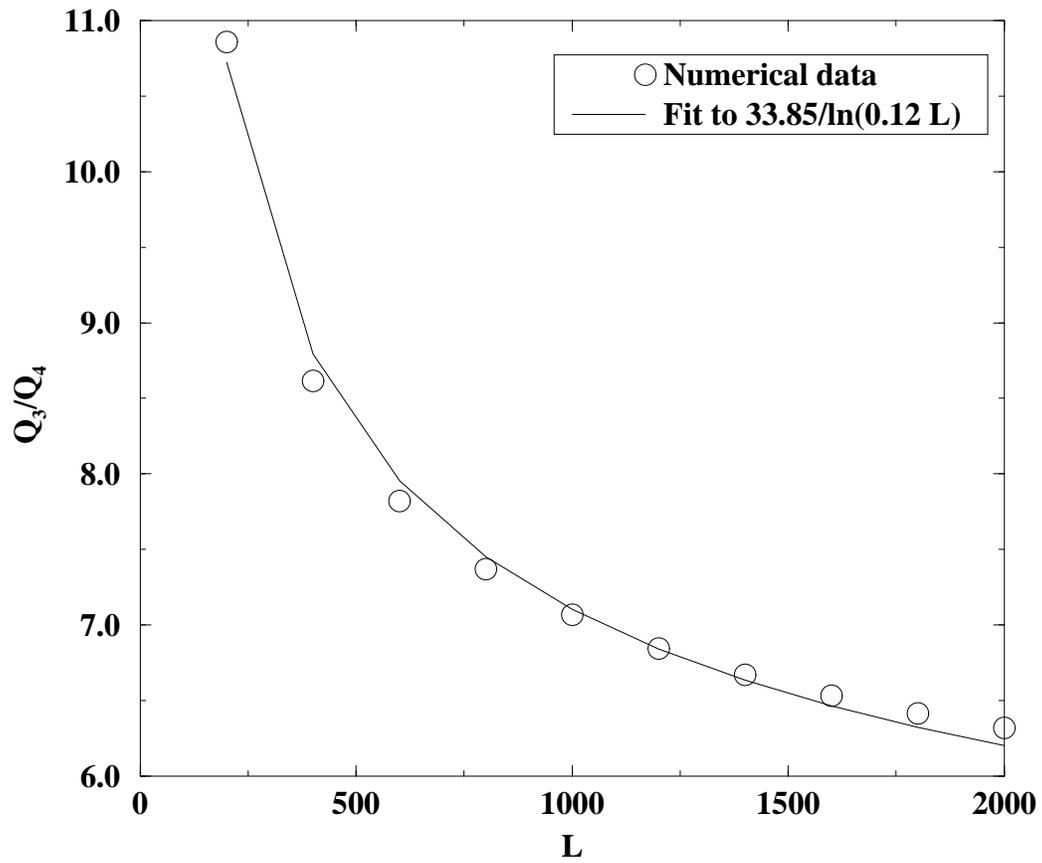}
\caption{Ratio test for the configuration shown in Fig.
\protect\ref{fi:conf_mf}. $Q_{a}$ is the average flux to particles of size
$ah$ (the average taken over the configuration). As shown by the figure,
the ratio is not independent of $L$, but includes a logarithmic factor
as discussed in the text.}
\label{fi:q3q4}
\end{figure}

\begin{figure}[t]
\psfig{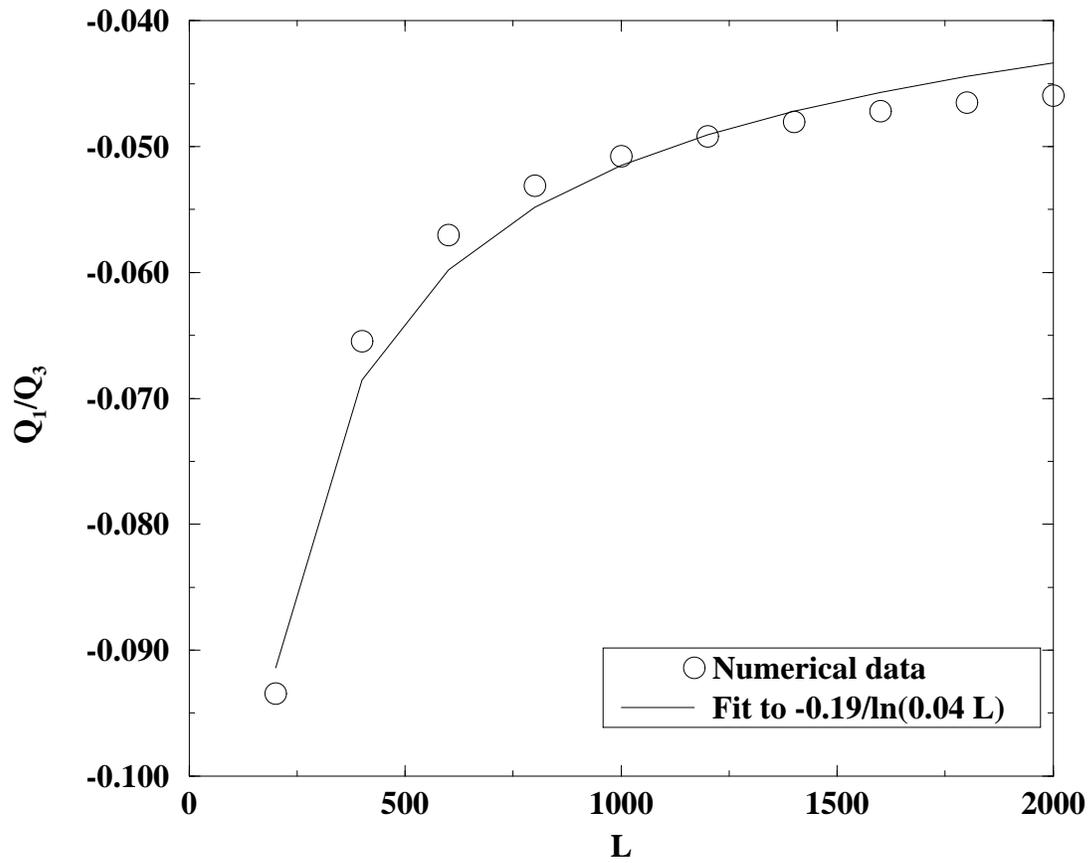}
\caption{Ratio test for the configuration shown in Fig.
\protect\ref{fi:conf_mf}. $Q_{a}$ is the average flux to particles of size
$ah$ (the average taken over the configuration).}
\label{fi:q1q3}
\end{figure}

Figures \ref{fi:q3q4} and \ref{fi:q1q3} show the results of our
investigation for the configuration of Fig. \ref{fi:conf_mf}. For these
cases, it is clear that the ratio test fails. In view of these results,
it seems very unlikely that there is a special class of configurations
for which the ratio of the expected values of the $Q$'s for any two
subset of particles (each with a specified set of $R_{i}$'s) could be
independent of scale. If this is so, then the reduced particle size
distribution in the 3D/2D system is not independent of scale and
self-similarity does not hold.

We summarize our principle results:
\begin{enumerate}
\item When the dimensionality of the precipitate particles and of the
diffusion matrix differ, self-similarity with one scaling length is not
possible as it violates mass conservation. Scaling with two or more
lengths is possible.
\item In the case of three dimensional particles embedded in a one
dimensional diffusion matrix, we present evidence of scaling with two
lengths;  the average particle size $\left< R \right>$, and the 
inter-particle spacing $\left< d \right> $. 
The growth laws are $\left< R \right> \propto t^{1/7}$,
$\left< d \right> \propto t^{3/7}$, and the distributions of the
corresponding reduced variables become independent of time.
\item If the distribution of reduced particle size $x = R / \left< R
\right>$ is independent of scale, we have shown that the expected growth
rate of particles of a given radius factors into a product of a function
of time only (i.e., $d \left< R \right> /dt$) and a function of $x$
only. It follows that the ratio of the expected growth rates of
particles of two different sizes, or of two groups of particles of
different sizes, is independent of scale. This ratio test becomes a
necessary condition for self-similarity.
\item We have presented evidence that coarsening of three dimensional
particles connected by a two dimensional diffusion matrix is not
self-similar. In particular, we have presented evidence from both mean
field arguments and from direct numerical solution of particles in a
Laplacian field, that the ratio test fails. The failure is weak
(logarithmic) and hence over short periods coarsening may appear
self-similar with $\left< R \right> \propto t^{1/4}$.
\end{enumerate}

\section*{Acknowledgments} The research of JV has been supported by the 
U.S. Department of Energy, contract No. DE-FG05-95ER14566, and also 
in part by the Supercomputer Computations Research Institute, which is
partially funded by the U.S. Department of Energy, contract No. 
DE-FC05-85ER25000. WWM is supported by the MRSEC Program of the National
Science Foundation, contract No. DMR-9632556.

\newpage
\appendix
\section{Expected rate of change of particle radius}
We start from the continuity equation for the particle radius distribution
\begin{equation}
\label{eq:apaa}
\left( \frac{ \partial n}{\partial t} \right)_{R} + 
\left( \frac{ \partial J}{\partial R} \right)_{t} = 0
\end{equation}
where $n(R,t)dR$ is the number of particles per unit \lq\lq area" in the
system and $J = n(R,t) u(R,t)$ in which $u$ is the average velocity of
particles along the $R$ axis, that is $u = \left< \dot{R} | R \right>$, or the
expected value of $dR/dt$ given the value of $R$.

Now assume that the particle distribution scales so that
\begin{equation}
\label{eq:apab}
n(R,t) = \frac{ N(t)}{\left< R \right> } P(x),
\end{equation}
where $P(x)$ with $x=R/ \left< R \right> $ is normalized on $x$;
thus expressed is a product of functions of $t$ only and $x$ only.
We seek the functional forms of $J$ and $u$.

To this end, we first convert Eq. (\ref{eq:apaa}) to express it in terms of $x$
and $t$. We have,
\begin{eqnarray}
\left( \frac{\partial n}{\partial t} \right)_{R} & = & \left( 
\frac{\partial n}{\partial t} \right)_{x} + \left( \frac{\partial n}{\partial
x} \right)_{t} \left( \frac{\partial x}{\partial t} \right)_{R} \nonumber\\
\label{eq:apabb}
& = & \left( \frac{\partial n}{\partial t} \right)_{x} - 
\frac{x}{\left< R \right> } \frac{d \left< R \right >}{dt} 
\left( \frac{\partial n}{\partial x} \right)_{t}
\end{eqnarray}
and
\begin{equation}
\label{eq:apad}
\left( \frac{ \partial J}{\partial R} \right)_{t} = 
\left( \frac{\partial J}{\partial R} \right)_{t}
\left( \frac{\partial x}{\partial R} \right)_{t} = 
\frac{1}{ \left< R \right> } \left( \frac{\partial J}{\partial x} \right)_{t}.
\end{equation}
Substituting these expressions into (\ref{eq:apaa}) we find,
\begin{equation}
\label{eq:apae}
\left( \frac{\partial J}{\partial x} \right)_{t} = x
\frac{d \left< R \right> }{dt} 
\left( \frac{\partial n}{\partial x} \right)_{t} -
\left< R \right> \left( \frac{\partial n}{\partial t} \right)_{x}
\end{equation}
as the equation of continuity in terms of $x$ and $t$. Now substitute
(\ref{eq:apabb}) into the right hand side of (\ref{eq:apae}) and then add and
subtract the term $(d \left< R \right> /dt)(N/ \left< R \right>)P$ which 
allows the derivatives to be combined with the result,
\begin{equation}
\label{eq:apaf}
\left( \frac{\partial J}{\partial x} \right)_{t} = 
\frac{d\left< R \right> }{dt} \frac{N}{\left< R \right> } \frac{d (xP)}{dx} 
- P \frac{dN}{dt}.
\end{equation}
But $N \left< R^{D_{p}} \right> = \alpha N \left< R \right>^{D_{p}}$ 
by virtue of (\ref{eq:apab}) where $\alpha$ is a constant. Therefore,
\begin{equation}
\label{eq:apag}
\frac{1}{N} \frac{dN}{dt} + \frac{D_{p}}{\left< R \right> } 
\frac{d \left< R \right> }{dt} = 0.
\end{equation}
Using this relation allows terms of (\ref{eq:apaf}) to be combined to yield
\begin{equation}
\label{eq:apah}
\left( \frac{ \partial J}{\partial x} \right)_{t} = - 
\dot{N} \left[ \frac{1}{D_{p}} \frac{d(xP)}{dx} + P \right]
\end{equation}
which shows that the left hand side is a product of functions of time and $x$
only. This expression may be integrated with the added arbitrary function of
time chose to cause $J$ to vanish at infinity. The result is,
\begin{equation}
\label{eq:apai}
J = n u = - \dot{N} \left[ \frac{x}{D_{p}} P - \int_{x}^{\infty} P(x') dx' 
\right]
\end{equation}
which shows that $J$ and hence $u$ are products of functions of $t$ only
and of $x$ only.

One can go farther by solving Eq. (\ref{eq:apai}) for $u$ and using
(\ref{eq:apabb}) again to get
\begin{equation}
\label{eq:apaj}
u(x,t) \equiv \left< \dot{R} | R \right> = \frac{d \left< R \right> }{dt} G(x)
\end{equation}
where
\begin{equation}
G(x) = x - \frac{D_{p}}{P} \int_{x}^{\infty} P(x') dx'.
\end{equation}
Therefore the decomposition (\ref{eq:aj}) or (\ref{eq:apaj})
is a necessary condition for the
scaling assumption of the particle size distribution.

\newpage
\bibliographystyle{prsty}
\bibliography{references}

\end{document}